\documentclass{sigchi}
\usepackage{times}
\usepackage{helvet}
\usepackage{courier}
\usepackage[draft]{hyperref}

\usepackage[utf8]{inputenc}
\usepackage{url}
\usepackage{microtype}
\usepackage{graphicx}
\usepackage{booktabs}
\usepackage{tabulary}
\usepackage{subcaption}
\usepackage{flushend}
\usepackage{amsmath}
\usepackage[backend=bibtex,natbib=true,style=sigchi,maxcitenames=1,maxbibnames=100,sortcites=true]{biblatex}

\renewcommand{\cite}{\citep}

\usepackage{enumitem}
\setlist{nolistsep}

\usepackage{algorithmicx}
\usepackage{algorithm}
\usepackage{algpseudocode}

\newfont{\mycrnotice}{ptmr8t at 7pt}
\newfont{\myconfname}{ptmri8t at 7pt}
\let\crnotice\mycrnotice%
\let\confname\myconfname%

\newcommand{\ie}{\emph{i.\,e.}}

\newcommand{\eg}{\emph{e.\,g.}}
 
\newcommand{\wrt}{with respect to }
\usepackage{gensymb}

\usepackage{ifthen}
\newboolean{inthesis}
\setboolean{inthesis}{false}

{\ifthenelse{ \boolean{inthesis} }{}{}
{\ifthenelse{ \boolean{inthesis} }{}{}
{\ifthenelse{ \boolean{inthesis} }{}{}
{\ifthenelse{ \boolean{inthesis} }{}{}
{\ifthenelse{ \boolean{inthesis} }{\newcommand{\spara}[1]{\subsection{#1}}}{\newcommand{\spara}[1]{\smallskip\noindent{\bf #1.}}}

\addtolength{\belowcaptionskip}{-5mm}

\usepackage{xpatch}

\xpatchbibmacro{name:andothers}{%
  \bibstring{andothers}%
}{%
  \bibstring[\emph]{andothers}%
}{}{}

\numberofauthors{3}

\author{
\alignauthor
Eduardo Graells-Garrido\thanks{Corresponding author: \url{eduardo.graells@telefonica.com}. Work carried out while the first author was a PhD student in the Web Research Group, at Universitat Pompeu Fabra, Barcelona, Spain.} \\
\affaddr{Telefónica I+D} \\
\affaddr{Santiago, Chile} 
\alignauthor
Mounia Lalmas \\
\affaddr{Yahoo Labs} \\
\affaddr{London, UK} 
\alignauthor
Ricardo Baeza-Yates \\
\affaddr{Yahoo Labs} \\
\affaddr{Sunnyvale, USA} 
}

\toappear{In review. Do not distribute. Contact the authors before citing.}

\title{Encouraging Diversity- and Representation-Awareness \\ in Geographically Centralized Content}

\clubpenalty=10000
\widowpenalty=10000

\addbibresource{geodiversity.bib}

\begin{document}
 
\maketitle

\begin{abstract}
In centralized countries, not only population, media and economic power are concentrated, but people give more attention to central locations.
While this is not inherently bad, this behavior extends to micro-blogging platforms: central locations get more attention in terms of information flow.
In this paper we study the effects of an information filtering algorithm that decentralizes content in such platforms.
Particularly, we find that users from non-central locations were not able to identify the geographical diversity on timelines generated by the algorithm, which were diverse by construction.
To make users see the inherent diversity, we define a design rationale to approach this problem, focused on an already known visualization technique: treemaps.
Using interaction data from an ``in the wild'' deployment of our proposed system, we find that, even though there are effects of centralization in exploratory user behavior, the treemap was able to make users see the inherent geographical diversity of timelines, and engage with user generated content.
With these results in mind, we propose practical actions for micro-blogging platforms to account for the differences and biased behavior induced by centralization.
\end{abstract}

\sloppy 


\section{Introduction}
In his book on user experience, Bill Buxton said that \textit{``in order to design a tool, we must make our best efforts to understand the larger social and physical context within which it is intended to function''} \cite{buxton2010sketching}. 
In today's global Web, it is not clear if current social platforms consider those different contexts when building their user interfaces or defining their content-based algorithms.
This would not be a problem in an uniform, unbiased world, but our world is neither uniform \cite{hofstede2010cultures} nor unbiased \cite{merton1996social}.
One of these biases is \textit{centralization}~\cite{kollman2013perils}, an organizational schema for governments that can be beneficial for the population, for development, and for the economy, because of the concentration of  political, economical and media powers.
However, at the same time, it can create regional inequalities within a country \cite{krugman1999role}, making national growth biased toward central locations.
Given that centralization is part of the social context of users, in this work we evaluate whether this bias affects the informational and exploratory behavior of users on the Web, particularly in micro-blogging platforms.
We do so through the following research questions:

\begin{quote}
Does centralization affect how people perceive information, and how people behave when browsing informational content in micro-blogging platforms? If so, how can we encourage non-centralized exploration?
\end{quote}

To answer our research questions, we analyzed the usage of Twitter by people in Chile interested in politics.
This case study is perfect for our analysis for two reasons. 
Chile is a highly centralized country toward its capital region~\cite{atienza2012concentracion,galiani2008political}.
Although this region is indeed near the geographical center of continental Chile, the country spans over 4,300 km (2,671 mi) from north to south, with only 175 km (108 mi) in average from east to west, having a geographical configuration that is not particularly fit for centralization.
After returning to democracy, various initiatives attempted to decentralize the country, but enormous political and economical resistance allowed for only small progress towards this, if any \cite{eaton2004designing}. 
On the other hand, Chile is one of the developing countries with the highest Internet penetration rate \cite{pewresearch2015}, and Twitter is actively used by politically-involved Chileans \cite{valenzuela2014facebook}. 

A previous study \cite{graellsbalancing} determined that centralization is reflected in the way users interact in Twitter, creating regional inequalities in information flow \wrt user location. The authors developed an information filtering algorithm that generates geographically diverse timelines. By construction, the algorithm allows all locations to be present in the timeline. 
In this paper, we evaluate this algorithm through a user study and found that users from central locations have a different sense of diversity, interestingness, and informativeness than users from peripheral locations. 

The results from the user study imply that, even though in theory the algorithm solves the problem, in practice users might not notice it.
After analyzing these differences, we hypothesize that centralization generates a \emph{diversity- and representation-awareness} problem, and thus we define a design rationale to address these issues. 
In compliance with our design rationale we then tested a visualization of categorical news headlines by \citet{newsmap} in our context, \ie, in geographically diverse timelines with micro-posts instead of headlines. We hypothesize that this interface eases the identification in timelines by making geographical diversity salient through the usage of \emph{squarified treemaps} \cite{bruls2000squarified}, a well-known information visualization technique.
We evaluated this design in an exploratory application deployed on the Web.
An analysis of interaction data revealed that, even though there are differences in how people behave \wrt their geographical origin, our design facilitates diversity- and representation-awareness from a geographical point of view.

Our results show that centralization affects the perception and behavior of politically-involved users in Chile; they also demonstrate that awareness of social and geographical contexts when designing algorithms and user interfaces improves user perception in the presence of these systemic biases. 
Finally, considering that many countries are centralized \cite{galiani2008political,kollman2013perils,krugman1999role}, our findings and implications can be used by micro-blogging platforms to make their systems aware of important factors of social contexts.

\section{State of the Art}

This paper deals with \textit{centralization} \cite{kollman2013perils}, in particular considering its geographical aspect.
Geography is an important attribute to be considered in the study of social networks and Web platforms. In the context of Twitter, 
even though most of user ties are geographically local \cite{quercia2012social}, more than a third of mentions and links are inter-countries \cite{kulshrestha2012geographic}.
To understand how a virtual population is distributed, each user's location must be determined, a meta-attribute not always available.
When geolocating users in micro-blogging platforms, a simple approach is to query a gazetteer with the user's self-reported location~\cite{hecht2011tweets}, 
but there are more complex and accurate approaches that involve entity recognition \cite{abel2012semantics} and language models \cite{cheng2010you}. 
In our context, location classifiers can become biased when the population is imbalanced \cite{rout2013s}, and care must be taken when parameterizing such algorithms.

Despite being a virtual platform, Twitter users reflect or are influenced by physical world phenomena.
For instance, the number of international flights between countries is the best predictor of non-local ties \cite{takhteyev2012geography};
tweets can predict gross community happiness by comparing aggregated word usage related to mood in communities \cite{quercia2012tracking};
interaction and publishing behavior on Twitter is related with cultural dimensions \cite{ICWSM136102};
and international communication is still dominated by physical distance \cite{garcia2014twitter}.
Communication and tie formation in Twitter are influenced by geography and culture---two aspects deeply connected with centralization.
In Chile, the informational behavior of the population is centralized \cite{graellsbalancing}.
Thus, in our work, we evaluate if centralization influences user perception of geographically diverse timelines.

We focus on geographical diversity because under population imbalance and centralization, it is expected that timeline content is also imbalanced, both in volume and in popularity metrics (\eg, number of retweets).
Then, we propose to filter timelines using methods that increase geographical diversity of its content.
Diversification can be performed by minimizing similarity between items in a recommendation list~\cite{ziegler2005improving}, maximizing entropy of a set of content features~\cite{de2011identifying}, as well as context-specific diversification methods~\cite{munson2009sidelines}.
In the case of Twitter, diversified timelines (in its information entropy sense) have been found to be easier to remember and more engaging~\cite{de2011identifying}. 
In this paper we build upon such work, by evaluating perception of timelines generated by an information filtering algorithm \cite{graellsbalancing}, which, in turn, is based on two previous algorithms that aim to generate diverse information sets \cite{munson2009sidelines,de2011identifying}.

We focus on how user location influences those perceptions and how engagement differs in terms of location and user interface.
\citet{park2009newscube} showed that the presentation of news headlines in clusters generates more clicks on news than non-clustered displays. 
We extend their design guidelines by using \textit{squarified treemaps}~\cite{bruls2000squarified} to maintain clustered representations of micro-posts while, at the same time, making diversity visually salient and noticeable. 
Treemaps have been used before to visualize content from micro-blogging platforms by \citet{archambault2011themecrowds}, although their approach is different to ours: they visualize clustered keywords, whereas we visualize entire tweets using a design inspired by \textit{Newsmap.jp} \cite{newsmap}, a treemap visualization of news headlines.
In the context of political diversity, it has been acknowledged that visualization-based interfaces make people behave differently \cite{faridani2010opinion}.

It is known that people from different cultures behave in different ways when communicating, not only on micro-blogs~\cite{garcia2014twitter} but also on other forms of communication, like instant messaging \cite{kayan2006cultural}. 
In our work, we propose that centralization introduce differences in how users behave, and thus, these differences should be accounted for when designing systems.

\section{Background}
We describe the dataset and methods used by \citet{graellsbalancing}, which provide the background of this work.
Although the defined methods are general, we restrict ourselves to Twitter, a micro-blogging platform where users publish status updates called \emph{tweets} with a maximum length of 140 characters.
Users can \emph{follow} other users, establishing directed connections between them.
When user \emph{A} \emph{follows} user \emph{B}, tweets and \emph{re-tweets} made by \emph{B} will show up in \emph{A}'s timeline.
A timeline is a list of tweets in reverse chronological order.
Users can annotate tweets using \emph{hashtags}, \ie, keywords that start with the hash character \texttt{\#}.

\spara{Context: Municipal Elections in Chile}
In the first user study on this paper we analyze a dataset of tweets from Chile. 
Centralization in Chile is characterized through geography at the \textit{regional level}, \ie, the first-level administrative division of the country.
Chile has 15 regions, and the capital \emph{Regi\'on Metropolitana} (translated as \textit{Metropolitan Region}, and \textit{RM} hereafter) is the most central one \cite{atienza2012concentracion,galiani2008political}. 

The dataset is composed of tweets crawled on October 28th, 2012, in the context of municipal elections held in Chile that day. 
The event had a distinctive hashtag (\textit{\#municipales2012}), which, among other related hashtags (e.g.~\textit{\#tudecides}), keywords (e.g.~\textit{vote}), location and candidate names, were used as queries for the \emph{Twitter Streaming API}.\footnote{\url{https://dev.twitter.com/docs/streaming-apis}} 
In total, this dataset contains 157,648 users who published 724,890 tweets, and the user distribution is geographically representative of the population.
Because its content is about local elections happening nationwide, this dataset is suitable to study the effects of centralization.
This was done in \cite{graellsbalancing} by aggregating \textit{mentions}, \textit{replies} and \textit{retweets}. 
According to the aggregated regions of origin and destination of these interactions, an interaction graph between locations is built, in which the authors estimated \emph{betweenness centrality} using random walks \cite{newman2005measure}.
Betweenness centrality was chosen instead of other centrality metrics because it represents \emph{``the potential of a point for control of information flow in the network''} \cite{freeman1977set}.

\begin{figure}[tbp]
 \centering
 \includegraphics[width=0.9\linewidth]{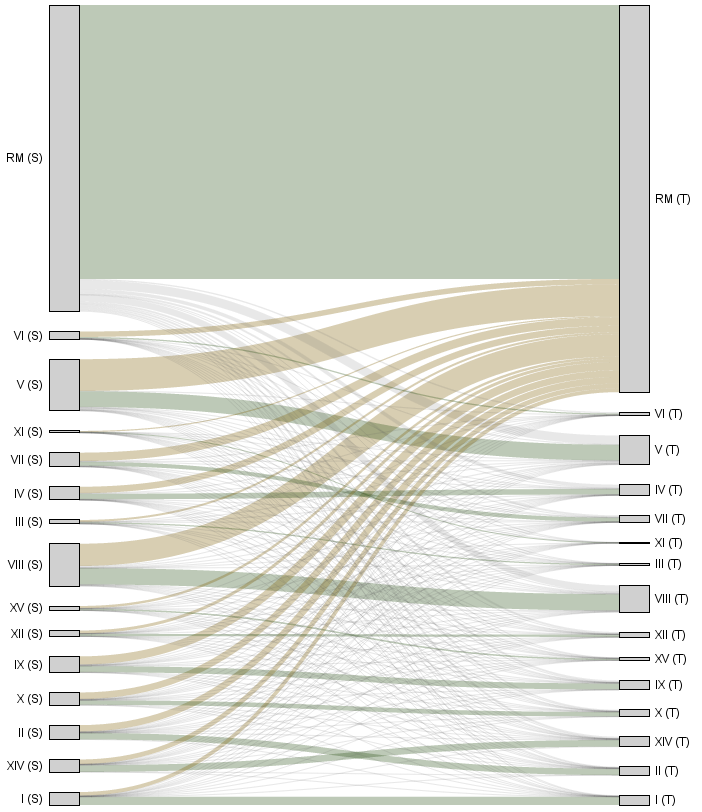}
\caption{Depiction of interactions between Chilean locations, using the dataset from \cite{graellsbalancing}. Source locations are on the left, and target locations are on the right. Edge color encodes interaction with self (\textit{green}), with \textit{RM} (\textit{beige}), and with others (\textit{gray}). Node order minimizes the number of crossings between edges.}
 \label{fig:chilean_centralization}
\end{figure}

The use of a random walk allowed the authors to consider the population distribution instead of binary edges in the graph, as well as to build a baseline graph based on the population distribution of the physical world. Then, by comparing both graphs, it was possible to determine whether informational behavior in Twitter, in terms of to whom people interact with, is centralized. It was found that \textit{1)} \emph{RM} is the most central location in Twitter discussion; \textit{2)} that centralities are significantly different between observations (interactions) and expectations (physical world population distribution); and \textit{3)} the observed centrality of \textit{RM} is higher than the expected one, and the centralities of all other locations exhibit the opposite behavior, implying that centralization also affects informational behavior in political discussion of Chileans in Twitter. A flow depiction of this graph is showed on Figure \ref{fig:chilean_centralization}. 

\spara{Building Geographically Diverse Timelines}
The influence of centralization makes content from peripheral locations harder to find.
One way to help to reduce this bias is through generation of geographically diverse timelines where all regions are represented regardless of their centrality.
This can be achieved through the usage of information filtering techniques.

\begin{algorithm}[tbp]
\caption{Geo. Diverse Information Filtering Algorithm.}
\label{alg:filtering_timeline}
\scriptsize
\begin{algorithmic}
\Require $T \gets \text{set of microposts to be filtered}$
\Require $s \gets \text{cardinality of resulting filtered set}$
\Require $\text{turns} \gets \text{number of turns for sidelining}$
\Ensure $T_{\theta} \gets \text{filtered micropost set}$

\Function{$\text{geodiverse\_filtering}$}{$(T, s, \text{turns})$}
\State $T_{\theta} \gets \text{list()}$
\State $\text{sidelined} \gets \text{dictionary}()$
\ForAll{$\ell \text{ in } L$}  
   \State $\text{sidelined}[\ell] \gets 0$
\EndFor

\State $t \gets \text{random.choice}(\text{most\_popular\_microposts}(T_{E}))$
\State $T_{\theta}.\text{append}(t)$
\State $\text{sidelined}[t_{\ell}] \gets \text{turns}$

\Repeat  
  \State $T_{c} \gets \text{list()}$
  \ForAll{$t \text{ in } T_{E} \text{ not in } T_{\theta}$}  
      \If{$\text{max\_entr}(t, T_{\theta}) \text{ and } \text{sidelined}[t_{\ell}] \leq 0$}
          \State $T_{c}.\text{append}(t)$
      \EndIf
   \EndFor
   
   \State $t \gets \text{random.choice}(\text{popular\_microposts}(T_{c}))$
   
   \State $T_{\theta}.\text{append}(t)$
   \State $\text{sidelined}[t_{\ell}] \gets \text{turns} + 1$
   
   \ForAll{$\ell \text{ in } L$}  
      \State $\text{sidelined}[\ell] \gets \text{sidelined}[\ell] - 1$
      
   \EndFor
\Until{$|T_{\theta}| = s$}
\State \Return $T_{\theta}$
\EndFunction
\end{algorithmic}
\end{algorithm}

In this paper, we use an information filtering algorithm defined by \citet{graellsbalancing}. 
This \textit{Proposed Method} (hereafter \textit{PM}) mixes two previously known algorithms:

\begin{enumerate}
 \item 
The first algorithm \cite{de2011identifying} is greedy and based on \emph{information entropy}.
Entropy is estimated \wrt several content-based features extracted from tweets (presence of links, topical information in hashtags, author connectivity and experience, age of content, and so forth).
At each iteration, the algorithm selects a tweet from a candidate pool that maximizes the current entropy of the filtered timeline. 
Note that \cite{graellsbalancing} modified the algorithm to consider popularity (number of retweets) as a feature, by preferring candidate tweets who are more popular than others. 

\item
The second algorithm is known as \textit{Sidelines} \cite{munson2009sidelines}. 
Because the complexity of the information features considered in \cite{de2011identifying} can be greater than those of geography (\eg, consider the number of hashtags against the number of locations), the entropy contribution of these dimensions can be higher than the entropy contribution of geography.
Thus, geographical diversity is enforced by \textit{sidelining}, \ie, when a tweet is selected for inclusion by the first algorithm, its location will not be considered in the following $n$ iterations.
\end{enumerate}

In summary, the mixed algorithm applies both base algorithms simultaneously: tweets are selected according to their entropy contributions, but only considering those from locations that have not been sidelined (see Algorithm \ref{alg:filtering_timeline}). 
With this schema a timeline of $s$ tweets can be built, with ensured information and geographical diversity. 
Whether these timelines are perceived as more diverse, interesting and informative than baseline timelines (including the first base algorithm \cite{de2011identifying}) is what we evaluate in the next section.

\section{User Perception of Diverse Timelines}
\label{sec:user_perception}
 We describe the user study performed to evaluate how users perceive timelines generated with the \textit{PM} algorithm \cite{graellsbalancing} in comparison with those of baseline conditions.
In the study, users compare three user-centered attributes: \textit{diversity}, \textit{interestingness} and \textit{informativeness}.
Users are grouped according to their geographical origin: a centralized location (\textit{RM} group) or a peripheral one (\textit{NOT-RM} group).

\spara{Conditions and Datasets}
In addition to the \emph{Proposed Method} (\textit{PM}), we consider the following baseline conditions:
\emph{Popularity Sampling} (\textit{POP} hereafter): we select the $s$ most popular tweets in terms of retweets;
and \emph{Diversity Filtering} (\textit{DIV} hereafter): an implementation of the first base algorithm that maximizes information entropy of each timeline \cite{de2011identifying}.
Since the different conditions require pairwise comparisons, we split the dataset \cite{graellsbalancing} in three:
\begin{enumerate}
 \item \textit{morning-noon}: 140,211 tweets published by 52,403 users between 10:00AM and 2:30PM.
 \item \textit{afternoon}: 180,824 tweets published by 63,388 users between 2:30PM and 9:00PM.
 \item \textit{night}: 401,029 tweets published by 106,942 users between 9:00PM and 2:00AM (next day).
\end{enumerate}

This division of time matches the local culture where lunch happens between 1:00PM and 2:30PM, and dinner around 9:00PM. 
For each dataset we built filtered timelines with \textit{PM}, \textit{DIV}, and \textit{POP}.
We excluded retweets as our focus is on stand-alone, source tweets.

\spara{Participants}
Participants were recruited using \textit{snowball sampling} in Twitter using open calls to volunteer in the study, which were retweeted by participants.
No compensation was offered. 
We recruited 125 participants. 
Of them, 81 were male, 41 were female and 3 opted to not say.
In terms of age, 1 was 18--19, 59 were 20--29, 54 were 30--39, 4 were 40--49, 1 was 50+ years old and 6 opted not to say. 
All participants were from Chile (87 in \textit{RM} and 38 in \textit{NOT-RM}). 
Participants experience with social networks was asked using a five-point Likert scale: those from \textit{RM} scored 3.73 ($\sigma^2 = 0.59$), and those from \textit{NOT-RM} scored 3.68 ($\sigma^2 = 0.58$).

\spara{Experimental Setup}
For each time dataset we generated three timelines (with $s = 30$ tweets) using \textit{PM}, \textit{DIV}, and \textit{POP}.
Timelines were displayed with a format that resembled Twitter user interface.


\spara{Procedure}
The study had a \textit{within-subjects} design.
First, participants were asked to fill a questionnaire about demographic information and other features such as Twitter usage.
Then, in at most three steps, users performed a series of comparisons between two timelines rendered side by side, each one generated by a different condition.
To avoid sequence effects, the order of pairwise comparisons (\textit{POP/PM}, \textit{POP/DIV}, \textit{DIV/PM}) and the order of time datasets (\textit{morning-noon}, \textit{afternoon}, \textit{night}) were randomized in both, the position on the screen (left or right) and the experimental step.
Hence, all the participants contributed to all conditions whenever possible, as some participants were expected to not do all comparisons being an on-line, volunteered study.
Additionally, not all participants were expected to do a full read of timelines, and thus we discarded comparisons where the total reading time of both timelines was less than one minute.

\spara{Task}
Participants were instructed to read the two timelines side by side, and then answer the following questions: 
\begin{enumerate}
 \item \textit{Which of the two timelines is more diverse?}
 \item \textit{Which of the two timelines is more interesting?}
 \item \textit{Which of the two timelines is more informative?}
 \item \textit{Optional: Please explain your answers. Add examples if needed.}
\end{enumerate}

Questions 1 to 3 had a seven-point Likert scale from -3 to 3, where -3 (or 3) means that the timeline on the left (or right) was perceived as more \textit{diverse}, \textit{interesting} or \textit{informative} than the other, and a value of 0 means that there was no perceived difference.
Question 1 asked for general diversity as we did not want to prime participants into thinking primarily about geographical diversity, although we explained that it should be considered in its widest sense, including geography, by adding the following subtitle to the question: ``\textit{Consider diversity in its widest sense (geographical, demographical, topical, temporal, etc)}.''
Question 4 presented a free-text form element.
After answering the four questions, a pause screen was shown for 15 seconds to allow participants to rest.

\spara{Statistical Model}
The aforementioned questions define three dependent variables: \textit{diversity}, \textit{interestingness} and \textit{informativeness}.
For each dependent variable we built the following statistical model (using \textit{R formula syntax}):
\[Y \sim C(\text{comparison}) \times C(\text{location}) ,\]
where $Y$ is an ordered response variable, $C(\text{comparison})$ is a dummy variable that encodes the specific pairwise comparisons performed in the study, and $C(\text{location})$ is a variable that encodes whether users are from \textit{RM} or not.
The operator $\times$ encodes the main effects plus the interaction between factors.\footnote{$A \times B = A + B + A * B.$}

Over this model we performed a \textit{generalized linear model} regression with a \textit{proportional odds model} \cite{mccullagh1980regression}.
This model is also known as \textit{ordered logistic regression}, and is used when modeling ordinal dependent variables.
It extends the logistic regression model by allowing more than two categories, considering the order of the responses, and not assuming equidistant items in the Likert scale.
If the statistical interaction was not found to be significant, then we performed another regression without the interaction terms (\ie, using the $+$ operator instead of $\times$).

\begin{figure}[tb]
 \centering

 \includegraphics[width=\linewidth]{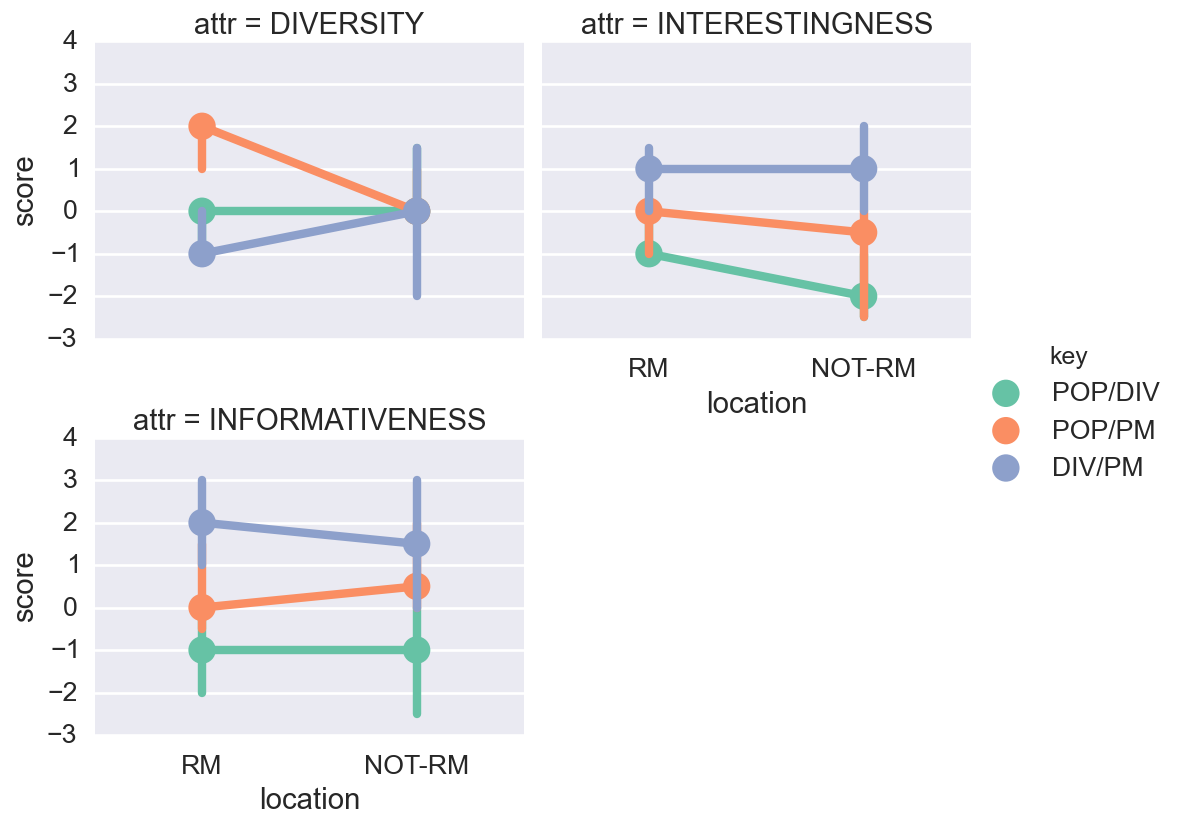}
 \caption{Point plots of median scores given by participants in comparisons of diversity, informativeness and interestingness.
For each pairwise comparison, a positive value indicates that the approach on the right of the label was perceived to be more diverse, interesting, and informative than the one on the left, and \emph{vice-versa}.}
 \label{fig:geodiversity_filtering_evaluation_violinplots}
\end{figure}

\begin{table*}[tb]
\centering
\footnotesize
\begin{tabulary}{\linewidth}{LLLCCCC}
\toprule
Result \# & Variable & Effect & $\beta$ & Odds-Ratio & 95\% C.I. & $p$ \\  
\midrule
R1 & Diversity       & Location \textit{RM} and condition \textit{POP/PM}    & $1.291$ & $3.637$ & $[0.317,  2.278]$ & $0.010$ \\
\midrule R2 & Interestingness  & Location RM & $0.686$ & $1.985$ & $[0.307, 1.069]$ & $< 0.001$ \\
R3 & Interestingness  & Comparison \textit{POP/DIV} & $-1.404$ & $0.246$ & $[-1.885, -0.933]$ & $< 0.001$ \\
R4 & Interestingness  & Comparison \textit{POP/PM} & $-0.859$ & $0.423$ & $[-1.353, -0.373]$ & $< 0.001$ \\
\midrule R5 & Informativeness      & Location \textit{RM} & $1.364$ & $3.910$ & $[0.869, 1.872]$ & $< 0.001$ \\
R6 & Informativeness      & Comparison \textit{POP/DIV} & $-0.926$ & $0.396$ & $[-1.618, -0.241]$ & $0.008$ \\
R7 & Informativeness      & Location \textit{RM} and comparison \textit{POP/DIV} & $-1.201$ & $0.301$ & $[-2.148, -0.261]$ & $0.012$ \\
R8 & Informativeness      & Location \textit{RM} and comparison \textit{POP/PM} & $-1.543$ & $0.214$ & $[-2.577, -0.521]$ & $0.003$ \\
\bottomrule
\end{tabulary}
\caption{Significant effects from the user study, including coefficients, effect sizes (Odds-Ratios) and confidence intervals. We only include effects with significant p-values ($p < 0.015$).}
\label{table:user_study_comparison_results}
\end{table*}

\subsection{Results}
In total, participants performed 238 comparisons: 84 for conditions \textit{POP/DIV}, 80 for conditions \textit{POP/PM}, and 74 for conditions \textit{DIV/PM}.
The following are the interpolated medians of pairwise comparisons given by users:

\begin{itemize}
 \item \textit{POP/DIV}: diversity (0), interestingness (-1), informativeness (-1). Users perceived both as equally diverse, but POP is found to be more interesting and informative.
 \item \textit{POP/PM}: diversity (1), interestingness (0), informativeness (0). PM is found by users to be more diverse than POP, and both are perceived as equally interesting and informative.
 \item \textit{DIV/PM}: diversity (-0.5), interestingness (1), informativeness (2). DIV is found by users to be slightly more diverse than PM, but PM is perceived as more interesting and informative.
\end{itemize}

Figure \ref{fig:geodiversity_filtering_evaluation_violinplots} showcases the medians of pairwise comparisons between conditions (questions 1, 2 and 3), considering individual differences according to location. It can be observed that statistical interactions could appear in the three attributes under consideration, specially with diversity.
To formalize this intuition, we applied the regression defined previously to the user scores of comparisons. The following list describes the fit of those regressions:

\begin{itemize}
 \item Diversity: the regression with interaction terms (log-likelihood $= -447.21$, AIC $= 910.41$, cond. H $= 1,900$) presented a significant interaction, and is used for analysis.
 \item Interestingness: the regression with interaction terms did not present significant terms according to the likelihood ratio test.
The regression without interaction (log-likelihood $= -430.84$, AIC $= 873.67$, cond. H $= 460$) is used for analysis.
 \item Informativeness: the regression with interaction terms (log-likelihood $= -429.35$, AIC $= 874.70$, cond. H $= 200$) presented a significant interaction, and is used for analysis.
\end{itemize}

The condition numbers of the respective Hessians indicate that the fits are not ill-defined.
Table \ref{table:user_study_comparison_results} shows the significant factors of each regression.
We identify each result as R$i$, to reference it later in the discussion.
We mention only those that are significant according to their p-value (considering $p \le 0.05$).

\spara{Qualitative Feedback}
In general, answers to those questions explain in detail the scores given by users to filtering algorithms in terms of interestingness and informativeness. For instance, users mentioned that \textit{DIV} was more ``noisy'' than \textit{PM} and \textit{POP}, and that \textit{POP} and \textit{PM} had more ``serious'' users (\eg, journalists or other popular accounts) who published data and facts about the elections, not just personal opinions. Both appreciations could be caused by the consideration of popularity in both algorithms. 
Other users valued \textit{PM} over \textit{POP} because \textit{RM} was not the only location being discussed.
However, this feedback did not explain the differences reported in diversity by users.

\begin{figure*}[!ht]
 \centering
 \includegraphics[width=0.95\linewidth]{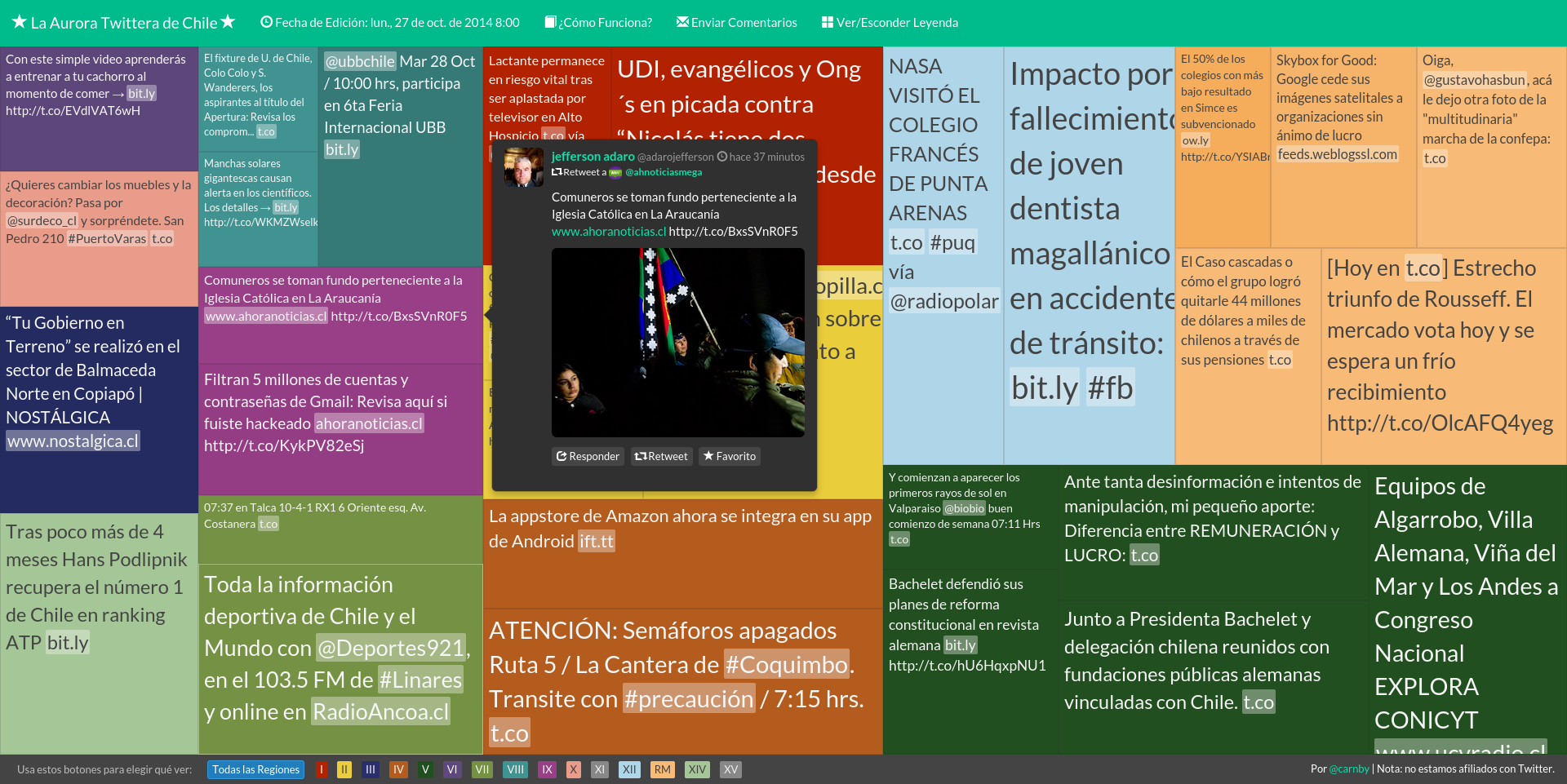}
 \caption{Screenshot from \url{http://auroratwittera.cl}, the URL of our prototype implementation. The bottom bar contains the location filters. The upper bar contain navigational links to an \textit{About} page and a feedback form.}
 \label{fig:aurora_virtual}
\end{figure*}

\subsection{Discussion of Results}
These results give us two important insights. 
First, \textit{popularity is slightly more valued than diversity}.
Users give more value to popularity than diversity, considering value as the mixture of informativeness and interestingness.
Even \textit{POP}, that is not geographically diverse, is perceived as more interesting than \textit{DIV} (R3) and PM (R4).
Likewise, \textit{POP} is perceived as more informative than \textit{DIV} (R6, R7 -\textit{RM} only-) and PM (R8 -\textit{RM} only-).
Yet, the Odd-Ratios (OR) of these results are low, and thus, while there is an effect, this effect is subtle.
Second, \textit{perception of timeline properties depends on the geographical origin of users}.
By design, we expected \textit{PM} to be perceived as more diverse by users than \textit{POP}, and more informative/interesting than \textit{DIV}.
However, this happened \emph{only when users come from \emph{RM}} (R1). 
Moreover, just being from \textit{RM} indicates that users are more prone to find content interesting/informative, regardless of its diversity (R2, R5). The high odd-ratios of these results indicate that the effect is strong.
The interaction between location and comparisons of \textit{POP} and the diversity-driven algorithms \textit{DIV} and \textit{PM} (R7, R8), indicates a slight (given its odd-ratios) accentuation of the effect.

From these insights, we hypothesize that people from central locations find \textit{PM} more diverse, and any approach more informative/interesting, because they are not used to see information from peripheral locations. 
It is known that the geographical span of ego-networks in Twitter is small \cite{quercia2012social}, and thus, exposing those users to views from other locations expands their vision.

In contrast, people from peripheral locations do not see differences in diversity, nor find the content informative/interesting, because they are used to be exposed to views from \textit{RM}.
For peripheral users, before filtering, the timeline content focused prominently on the centralized and most populated locations; after, it contained a wider set of locations, but still not prominently their own: \textit{``they are alike, we are diverse''}~\cite{quattrone1980perception}.
We recall that the source dataset guarantees the presence of local content, given that it is about nationwide municipal elections happening at the same time. This implies that users did not recognize the presence of their locations in the generated timelines.
Hence, in the next section we seek to enhance \textit{representation-awareness} of users, to help users to recognize their presence in the timeline.

\section{A Diversity-Aware Platform}
We propose a platform design that seeks to increase awareness of representation caused by diversity in timelines generated by the \textit{PM} algorithm. 
The platform is deployed ``in the wild'' \cite{crabtree2013introduction}, \ie, we target end-users in their everyday use of Twitter.

\subsection{Design Rationale: Awareness Through Identification}
Since \textit{PM} timelines are geographically diverse by definition, perhaps what people from peripheral locations need is to \emph{become aware} of their representation in the information stream. 
We propose to increase this awareness by facilitating \textit{identification} as defined by \citet{butler2006precarious}, \ie, identification relies upon differences with the others.
In our context, this means that by emphasizing differences with the others, we can facilitate identification (and thus, representation-awareness).
We do so by making diversity salient, allowing users to \textit{see} the geographical diversity present in the timeline.

\spara{Making Diversity Salient}
Prior work determined that presentation~\cite{park2009newscube} and visualization~\cite{faridani2010opinion} change the way users behave in the presence of diverse information. 
In particular, the grouping of news headlines according to agreement with political positions improved user access to diverse information, measured in clicks on those headlines \cite{park2009newscube}.
Then, clustering content in locations would make easier for users to quickly see their own locations represented.
However, it must be applied with care because our context is different. 
In political contexts, information is usually classified into a bipartite separation of groups (\eg, \textit{democrats} and \textit{republicans}, \textit{conservative} and \textit{liberal}, etc.), whereas in our case the number of locations is larger (\ie, 15 Chilean regions), creating the need to scroll on the screen and thus inducing a positional bias, by giving more importance to those clusters already visible without scrolling. 

To avoid scrolling and its associated positional bias, we consider a previous visualization of news headlines by Weskamp~\cite{newsmap}, which uses a 2D space-filling layout algorithm to partition the available screen space: the \textit{squarified treemap layout} \cite{bruls2000squarified}.
We visually encode locations as \textit{internal nodes} and tweets as \textit{leaves}. 
Sibling leaves appear together.
The \textit{area size} of each leaf depends on the number of retweets of the corresponding tweet, and the number of followers and friends of its author, in inverse proportions to population location.
In this way, screen space is shared in a fair way between locations, as shown in Figure~\ref{fig:aurora_virtual}.
Each leaf node is colored according to its location (\textit{hue}) and its recency (\textit{saturation}). Internal nodes are not displayed  as they are not needed.

\spara{User Interaction}
To interact with the visualization, users can \textit{click} a leaf node to display a pop-up with detailed information about the corresponding tweet, with buttons to perform core Twitter interactions (reply, retweet, mark as favorite, follow) and a text format that resembles the typical tweet presentation. 
To \textit{filter locations}, for each location we display a button that, when clicked, updates the treemap to display only tweets originated from the selected location. 

\subsection{Prototype}
We implemented a prototype of the user interface (see Figure~\ref{fig:aurora_virtual}) using the \textit{d3.js} \cite{bostock2011d3} library.
This interface is named \textit{``Aurora Twittera de Chile''} (\textit{AT} hereafter) and is available at \url{http://auroratwittera.cl}. 
Every $30$ minutes a \emph{``new issue of AT''} was generated by the filtering algorithm (having $s = 30$ as size for each timeline and $n = 5$ for turns in the sideline step in \textit{PM}). 

We used the same implementation of the filtering algorithm \textit{PM} from the first user study, with two differences: first, for performance reasons, we did not use a location classifier for all tweets in the dataset. Instead, we considered only tweets from accounts with a known self-reported location, although we did consider non-geolocated tweets retweeted by those accounts.
Second, we avoided repeated authors or tweet content in the same timeline, and we discarded tweets where almost all text was in uppercase letters to avoid \textit{shouting}.

Input tweets were downloaded with a crawler using the \textit{Twitter Streaming API}. As query keywords we used location names, political terms, and other terms of interest that appear constantly on the news, as well as mentions to media accounts, both at national and local~levels. 
Each issue had a specific URL in the form \url{http://auroratwittera.cl/timeline/ID}, which allowed users to access it at a later time, as well as saving a permanent link.
In addition, if the URL contained the code of a location (\eg, \url{http://auroratwittera.cl/\#RM}) the interface displayed immediately the tweets related to that location in the same way as if a location filter button had been pressed.

\spara{Social Bot @todocl}
Since social bots can generate social discussion and behavioral changes based on their activity~\cite{aiello2012people}, 
we created a social bot on Twitter, with username \textit{@todocl}, to publicize \textit{AT} and recruit users.
\textit{@todocl} presented itself as a social experiment to establish an informative community about current happenings in Chile using Twitter, and published three types of tweets:

\begin{itemize}
 \item Whenever a timeline was generated, \textit{@todocl} published two tweets with a link to its corresponding \textit{issue}: one mentioning four users who authored ``featured tweets'', and one mentioning four users with ``featured retweets'' (in both cases users were selected randomly from the pool of tweets).
 \item After publishing those tweets, every minute \textit{@todocl} retweeted one tweet featured in the current issue.
 \item Every hour past 45 minutes, \textit{@todocl} published 15 tweets, one per location, featuring a link to the current issue with each  specific location in the URL, as well as an attached image with a \textit{wordcloud} of their most representative terms obtained with \textit{Term Frequency/Inverse Document Frequency} (TF-IDF).
\end{itemize}

This implementation, comprised of filtering algorithm, user interface and social bot, creates a platform where users can access geographically diverse information, in the form of an external application to Twitter, as well as injected into the platform itself.
We evaluate this application next.

\section{Evaluation with Interaction Data}
In this second study, we evaluated user behavior based on the interaction data obtained from \textit{AT}, in particular at differences \wrt centralization.
Note that \textit{AT} is not a task-based system, \ie, we do not expect users to visit the site to perform a specific task.
Instead, the system is designed as an exploratory interface to geographically diverse timelines. 
By being exploratory, common evaluation metrics such as task accuracy or task performance cannot be applied.
Instead, we evaluate user interaction with our system from the point of view of three dimensions: diversity-awareness, representation-awareness, and interestingness.
From the analysis of those dimensions, we elaborate plausible explanations of potential differences in behavior as indicated by statistical analysis.

\subsection{Experimental Setup}
In this experiment we analyze interaction data obtained from user interaction with our implementation in \textit{AT}.
Each interaction is considered an event. We logged implicit events on the server, such as \textit{session created/restored}, and from the client browser using Javascript, such as \textit{timeline and UI loaded}, and \textit{pings}.
As explicit events we logged clicks on every element of the user interface, allowing us to identify whether users filtered locations, clicked on links, asked for a more detailed view of a tweet, as well as clicking on the Twitter native buttons to reply, mark as favorite, or follow a tweet's author.

Using the logged events we analyze user behavior from three dimensions:

\begin{itemize}
 \item \textit{Diversity-Awareness}: defined as the number of different locations that generated the content the user interacted with. Every time the user interacted with content, we logged the location which originated it. Thus, we define a variable named \textit{distinct locations} that encodes the number of different locations the user has interacted with.
 \item \textit{Representation-Awareness}: defined as the likelihood of selecting specific locations for browsing by users. We define a variable named \textit{filter likelihood} that is 0 if the user has not selected specific locations using the location filters from the UI, and 1 otherwise.
 \item \textit{Interestingness}: defined as the number of interactions with content by the user per day. We define a variable named \textit{content events} that sums the number of clicks in links, the number of tweets seen with more detail, the number of tweets marked as favorite or replied, as well as the number of times other users have been followed. Given that some users visited the site more than once, we normalize this variable according to the number of different days each user has accessed the site.
\end{itemize}

\spara{Participants}
We gathered interaction data from October 1st, 2014 until January 20th, 2015. 
The server logged each user request and was able to identify sessions based on cookies placed on user browsers.
IP addresses were used to identify each user's location, using the GeoIP Legacy Database.\footnote{\url{http://dev.maxmind.com/geoip/legacy/geolite/}}
The User Agent information was used to determine if the user was browsing from a mobile device; those users were served with a minimal version of the site but were not considered in the study because of platform heterogeneity (in terms of interaction capabilities, screen sizes, etc.).
From 16,969 valid interaction events, we have 321 users, of which 193 are in \textit{RM} and 128 are in \textit{NOT-RM}.
We discarded users who were active in the site for less than 10 seconds or that were in the top 5\% of dwell time,\footnote{Those users presumably left their browser windows open.} as well as those who deactivated Javascript or could not be geolocated using GeoIP.

\begin{figure}[tb]
 \centering
 \begin{subfigure}[b]{0.475\linewidth}
 \centering
        \includegraphics[width=\linewidth]{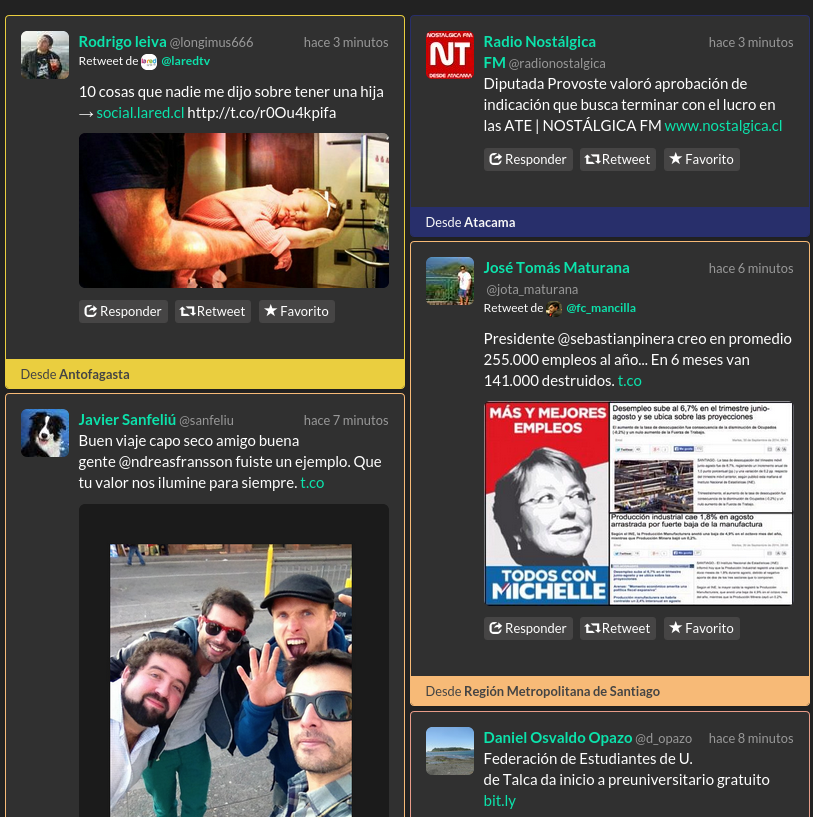}
\end{subfigure} 
 \begin{subfigure}[b]{0.475\linewidth}
 \centering
        \includegraphics[width=\linewidth]{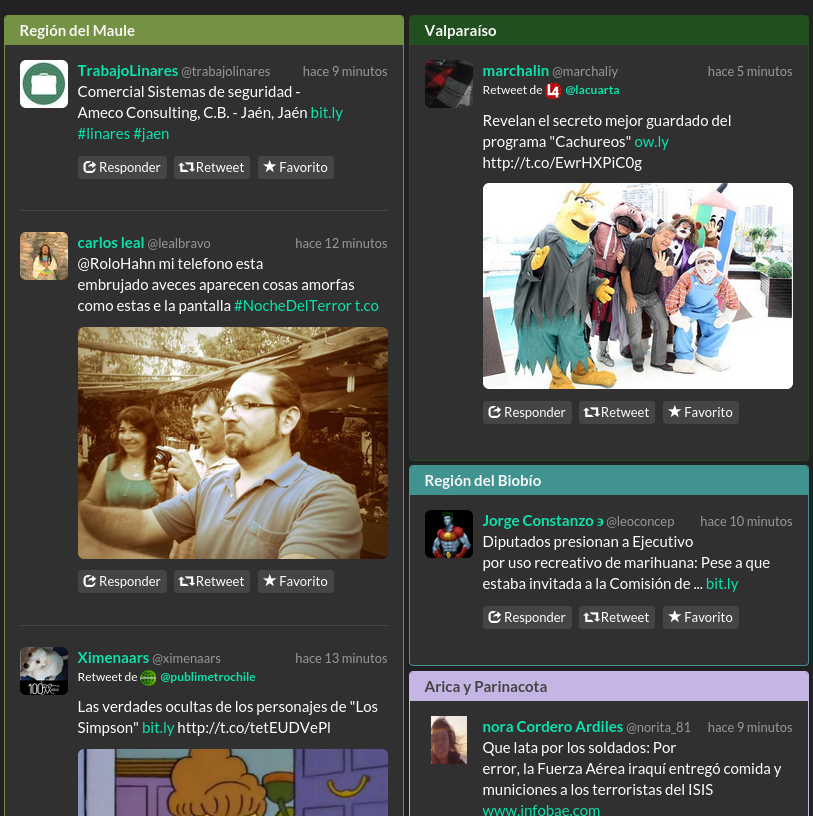}
\end{subfigure}%
 \caption{Design baselines implemented for the study. Left: standalone tweets (each in its own box). Right: clustered tweets by location.}
 \label{fig:design_baselines}
\end{figure}

\spara{Conditions}
To evaluate the effect of both, location and user interface, on user behavior, we developed two alternative baseline conditions to compare with our own design:
\textit{1)} \textit{Baseline}, where each tweet is rendered independently of the others, and tweets are sorted by their time of publication. Each tweet is displayed inside a box with a bordered color and a legend at the bottom to indicate its originating location; and
\textit{2)}~\textit{Clustered}~\cite{park2009newscube}, where tweets are clustered by location. Each location is represented as a box with a bordered color and a legend at the top to indicate the originating location of its tweets.
Both baselines are shown in Figure \ref{fig:design_baselines}. It can be seen that, although they are text-based, they include visual cues that indicate differences in location. Also, both have the same interactive location filters as our condition, \textit{Treemap}.

When a user accessed \textit{AT}, if it was his/her first visit, a random condition was assigned.
Because we tracked users using cookies, in following requests users received the same interface according to their initial assignment.
The distribution of users is as follows: \textit{baseline} was assigned to 97 users; \textit{clustered} to 98 users; and \textit{treemap} to 126 users.

\spara{Setup}
We consider a \textit{between-subjects} design, as participants were exposed to one condition only.
When users loaded \textit{AT} we gathered the following user information: \textit{IP address}, \textit{HTTP Referrer} and \textit{User Agent}. 
Then, we logged each interaction with elements of the user interface.
If the user requested the page through an URL with a location code, we considered it as an initial click on a location filter, because such links were generated by the social bot, and the user explicitly chose to follow them instead of the main URL.
Finally, the user interface sent a ping every ten seconds to the server to track the time spent on the website even in the absence of interactions, to capture the dwell time of passive users.

\spara{Statistical Model}
For analysis, we evaluate the effect of location and condition in the dependent variables defined earlier: \textit{distinct locations}, \textit{filter likelihood} and \textit{content events}.
We consider the two categorical independent variables \textit{location} (\textit{RM} or \textit{NOT-RM}) and \textit{condition} (\textit{baseline}, \textit{clustered}, or \textit{treemap}). 
Both are included in the following statistical model:
\[Y \sim C(\text{condition}) \times C(\text{location}) .\]
As in the previous study, we consider main effects of each factor as well as the interaction between them. The $C$ element in the formula generates \textit{contrasts} as dummy variables.

Over this model we perform several \textit{generalized linear model} regressions.
The link function used in each regression varies according to the meaning of each variable: 
for \textit{distinct locations} and \textit{content events} we use the \textit{Negative Binomial} distribution, which is commonly used to model over-dispersed count data;
and for \textit{filter likelihood} we use the \textit{logistic} (\textit{logit}) function, because we model a probability.
For each variable, if the statistical interaction was not found to be significant, we performed another regression without the interaction term.

\begin{figure}[!htb]
 \centering
 \includegraphics[width=\linewidth]{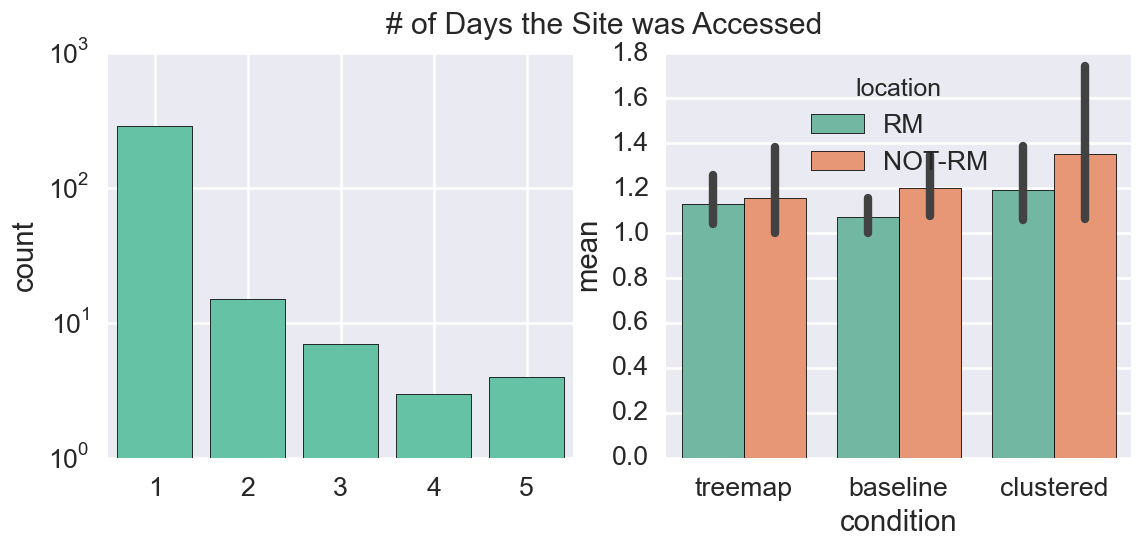}
 \includegraphics[width=\linewidth]{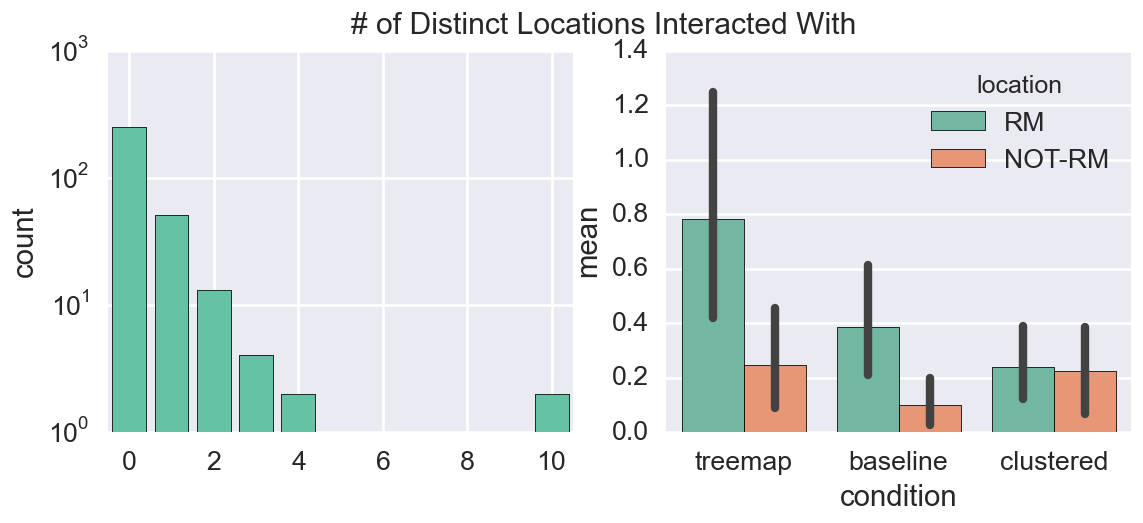}
 \includegraphics[width=\linewidth]{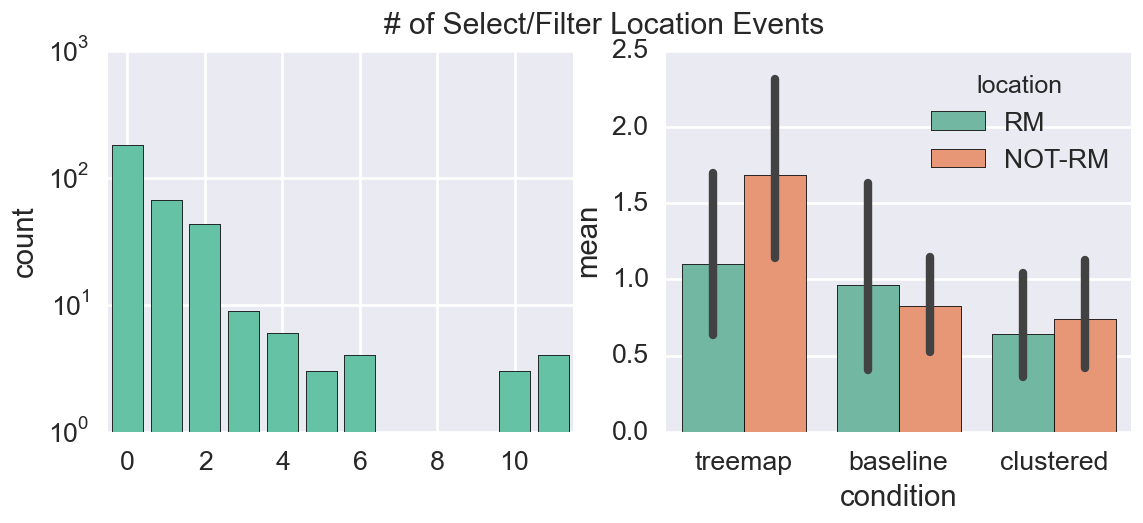}
 \includegraphics[width=\linewidth]{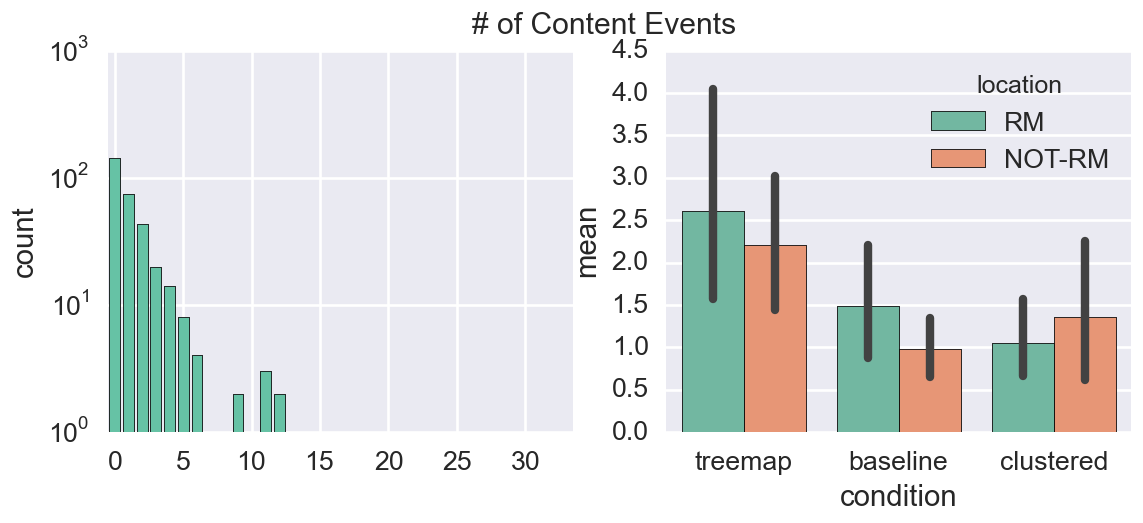}
 \caption{Distribution of each variable analyzed from interaction data. For each variable we include a histogram (left) and a bar plot (right) to compare means between groups.}
 \label{fig:variable_distributions}
\end{figure}

\begin{table*}[tbp]
\centering
\footnotesize
\begin{tabulary}{\linewidth}{LLLCCCC}
\toprule
Result \# & Variable & Effect & $\beta$ & O.R. & 95\% C.I. & $p$ \\  
\midrule
R9 & distinct\_locations (N.B.) & Condition \textit{treemap} & $0.732$ & $2.080$ & $[0.092, 1.373]$ & $0.025$ \\ 
R10 & distinct\_locations (N.B.) & Location \textit{RM} & $0.950$ & $2.473$ & $[0.348, 1.553]$ & $0.002$ \\ 
R11 & filter\_likelihood (logit) & Condition \textit{treemap} & $0.7541$ & $2.126$ & $[0.207, 1.301]$ & $0.007$ \\
R12 & filter\_likelihood (logit) & Location \textit{RM} & $-0.494$ & $0.610$ & $[-0.955, -0.033]$ & $0.036$ \\
R13 & content\_events (N.B.) & Condition \textit{treemap} & $0.628$ & $1.873$ & $[0.167, 1.088]$ & $0.008$ \\ 
\bottomrule
\end{tabulary}
\caption{Significant effects from the user study, including coefficients, odds-ratios, and confidence intervals.}
\label{table:user_interaction_regression_results}
\end{table*}

\subsection{Results}
Figure \ref{fig:variable_distributions} shows the distributions of our variables.
Additionally, we include the distribution of the number of days each user has visited the site, because it is used to normalize the variable content events, which is not normalized in the figure.\footnote{We applied a N.B. regression to the number of days, but no effect of condition/location was found.}
The following list describes the regression of each variable:

\begin{itemize}
 \item Variable distinct\_locations (mean $= 0.364$, std $= 1.013$, max $= 10$): 
 the \textit{N.B.} regression (deviance $= 248.44$, $\chi^2 = 453$, log-likelihood~$= -240.45$) has intercept $\beta~=-1.987$ (95\% C.I.  $[-2.683, -1.290]$, $p < 0.001$).
 \item Variable filter\_likelihood (mean $= 0.436$, std $= 0.497$, max $= 1$):
 the \textit{logit} regression (log-likelihood~$= -212.41$, $p = 0.002$) has non-significant intercept $\beta~=-0.289$ (95\% C.I.  $[-0.779, 0.201]$, $p = 0.247$).
 \item Variable content\_events (mean $= 1.688$, std $= 3.313$, max $= 33$):
 the \textit{N.B.} regression (deviance $= 374.79$, $\chi^2 = 564$, log-likelihood~$= -518.79$) has non-significant intercept $\beta~=0.105$ (95\% C.I.  $[-0.328, 0.538]$, $p = 0.635$).
\end{itemize}

Table \ref{table:user_interaction_regression_results} shows the significant coefficients of each regression.
We show only those with a p-value lesser than $0.05$.
We identify each result as R$i$, to reference it later in discussion. As effect sizes we analyze Odd-Ratios (OR).
In contrast with our first study, no interaction terms were significant.

\subsection{Discussion of Results}
\spara{Diversity- and Representation-Awareness}
With respect to diversity-awareness, we found that people from \textit{RM} are more likely to be diverse-aware (R10, OR $= 2.080$), which is coherent with the results from the first user study.
Moreover, we found that people exposed to the treemap condition were as likely to be diverse-aware (R9, OR $= 2.473$), which means that people from \textit{NOT-RM} exposed to the treemap can be as diversity-aware as people from \textit{RM}.

With respect to representation-awareness, we found that people from RM is less likely to select specific locations (R12, OR = 0.60), which means that people from NOT-RM is more likely to do so.
Moreover, we found that people exposed to the treemap condition were more likely to select specific locations (R11, OR = 2.126).

\spara{Interestingness}
We did not find an effect of location on the number of content events, which means that users from all locations were equally engaged  with content.
However, we found that people exposed to the treemap condition were more likely to interact with content (R13, OR $= 1.873$). 
Its OR indicates that using the treemap increases the chances of interaction with content by 87.3\%.

Our research question asks: \textit{how can we encourage non-centralized exploration?}
As interpreted from our first user study, it should never be just about the algorithm, but also about how users respond to what the algorithm returns to them.
This response is influenced not only by the algorithm, but also by the user interface used.
At the beginning of this section we stated that the main purpose of \textit{AT}'s design was to make users more diversity- and representation-aware.
We believe we succeeded, because, regardless of geographical origin (\ie, no statistical interaction), the treemap design showed good properties on the three dimensions under consideration. 
These properties, in terms of diversity, representation, and user interest, indicate that a good way to encourage such behavior is by allowing users to explore content using non-traditional user interfaces.
In fact, the clustered condition did not have any significant effect, even though it was expected~\cite{park2009newscube}. As outlined in our rationale, the number of clusters can make exploration more difficult for users.

\section{Implications}
In the literature, cultural differences \cite{hofstede2010cultures} have been acknowledged in the study of communication systems \cite{kayan2006cultural} by suggesting specific features and interaction mechanisms pertinent to each culture. 
As we have found in our studies, Chilean information seekers and content explorers in Twitter are affected by centralization. 
Their perception of content is different, and their interaction with exploratory interfaces is also different. 

This implies that location \wrt centralization (\ie, central or peripheral) introduces an individual difference to be accounted for in system design. 
If this difference is not considered, user exploration can be biased toward central locations. 
While centralized exploration is not inherently bad (\ie, there could be genuine interest in central location content given contingency or current news), the purpose of  decentralization is to make users aware of the geographical diversity of the information space.
What we do on-line, influences our off-line lives, and thus, such awareness would help users in today's \textit{``culture of real virtuality''} \cite{castells2011rise} to reaffirm their identities on the Web by recognizing their local traits and culture on content. 

To do so in exploratory settings, we propose the following design changes in micro-blogging platforms:

\begin{enumerate}
 \item Search results are ranked (presumably) by relevance and popularity. Centralized contexts (such as many Latin American countries) will have biased popularity metrics toward central locations. A diversity-aware algorithm (\eg, \textit{PM}) would help in this case.
 \item Currently, Twitter offers two options related to geography in its search page: ``Everywhere'' (the default) and ``Near me''. We suggest to change the default value according to the user's context, \ie, use ``Near me'' when users are from peripheral locations.
 \item Since including geographical diversity in timelines is not enough, because users may not be aware of it, techniques to make them \textit{see} this difference can help.
 We recommend platforms to use visualization techniques such as treemaps to display timelines. 
 The original design by \citet{newsmap} has proven to be useful when visualizing news headlines as it \textit{``allows many interesting comparisons and readings of how we differ culturally''} \cite{meirelles2013design}. Our results provide evidence of the good properties of treemaps, in particular in displaying the differences derived from user location and its relation to centralization.
\end{enumerate}

By considering these implications and suggestions, micro-blogging platforms will allow users to \textit{decentralize} part of their exploratory behavior.
Or, they could keep a centralized exploration in a \textit{conscious} way. The important outcome is that users will be able to choose \textit{how} and \textit{what} to explore.

\section{Conclusions}
In his essay \textit{``In praise of shadows''}, Junichir{\~{o}} Tanizaki wonders what if the fountain pen, an \textit{``insignificant little piece of writing equipment''}, would have been invented in Japan, as in spite of its insignificance it \textit{``had a vast, almost boundless, influence on our culture''}. 
In that line of thinking, we wonder ``\emph{what if global web platforms would have emerged in countries with severe systemic biases?}''
Perhaps algorithms would have considered those biases and user interfaces would have been adapted to mitigate their effects.
To this end, we studied the specific case of Chile, a highly centralized country \cite{galiani2008political}.
With Chilean users, considering their geographical origin from a centralization point of view, we analyzed through carefully designed experiments the differences in users' perception of diversity when exploring geographically diverse timelines.

In a user study with labeling tasks, we found that centralization induces differences in perception of what is interesting, what is informative, and what is diverse. 
In particular, only users from the centralized location were able to identify the geographical diversity present in the constructed timelines. 
Inspired by a visual design by \citet{newsmap}, we used information visualization to make users \emph{aware} of such diversity, addressing what we called the \textit{diversity- and representation-awareness} problem.
We deployed this design on the Web to display timelines generated with a diversity-aware algorithm, and spread information about it using the social bot \textit{@todocl}. 

By analyzing logged interaction data, we observed that users behave differently according to their geographical origin. We identified different interaction signals in central and peripheral users. Furthermore, we also observed that the proposed design improves exploration and diversity- and representation-awareness, regardless of geographical origin of users.
These results were used to define design implications and suggestions for micro-blogging platforms. 
We believe that consideration of our implications will have noticeable and positive consequences on information access by users in centralized contexts, which are not uncommon in Latin America \cite{galiani2008political}.

As of September 2015, \textit{@todocl} has more than 5,000 followers. We see this is as an indication of the need to provide content aggregators like \textit{AT}, that focus on diversification of specific social-context related aspects \cite{buxton2010sketching}.

\spara{Limitations and Future Work}
Critics might rightly say that our experiments suffer from biased user sampling and the lack of counterbalanced designs.
Currently, given centralization and the unequal population distribution, finding users from \textit{NOT-RM} is hard, because \textit{RM's} population is greater and has more access to the Internet in comparison to other locations, making it more difficult to find users willing to participate from non-central locations. 
Our snowball sampling method provided a way to find a large enough population to gain important insights in the first study.
In the second study, all users mentioned by \textit{@todocl} were geographically diverse, and those users retweeted \textit{@todocl}'s tweets, improving the representativity of the sample.
In terms of design, we could not validate users until the end of each study. 
Thus, because several users would have been discarded, the ideal counterbalanced designs would have proven to be unbalanced in practice.

The Web is increasingly being more accessed from mobile devices than from desktops.
In our case, of 1335 detected mobile users, 68\% could be geolocated using the IP address.
Considering those users in a new mobile-friendly version of our site will allow us to study differences in a mobile context.
Finally, we will study response to the social bot \textit{@todocl}, as we noticed that some users started to interact with it. This interaction could be compared with user engagement with traditional media accounts.

\spara{Acknowledgments}
We thank Pajarito, Denis Parra, Luz Rello, and Sergio Salgado for their helpful comments.
This work was partially funded by Grant TIN2012-38741 (Understanding Social Media: An Integrated Data Mining Approach) of the Ministry of Economy and Competitiveness of Spain.

\printbibliography

\end{document}